\documentclass[prd,aps,twocolumn,nofootinbib,superscriptaddress,eqsecnum,floatfix,preprintnumbers,amsmath,amssymb,
nofootinbib,longbibliography]{revtex4-2}

\usepackage{amssymb,amsmath}
\usepackage{epsfig}
\usepackage[dvipsnames]{xcolor}
\usepackage[utf8]{inputenc}
\usepackage{stmaryrd}
\usepackage{mathrsfs}
\usepackage{mathalfa}
\usepackage{accents}
\usepackage[normalem]{ulem}
\usepackage{enumitem}

\usepackage{graphicx}
\usepackage{textcomp}

\newcommand{\pa}{\partial}

\newcommand{\be}{\begin{equation}}
\newcommand{\ee}{\end{equation}}
\newcommand{\bea}{\begin{eqnarray}}
\newcommand{\eea}{\end{eqnarray}}
\newcommand{\ba}{\begin{equation}\begin{aligned}}
\newcommand{\ea}{\end{aligned}\end{equation}}

\newcommand{\beg}{\begin{gather*}}
\newcommand{\eng}{\end{gather*}}
\newcommand{\hh}{,\hspace{0.5cm}}
\newcommand{\hhh}{,\hspace{0.2cm}}

\newcommand{\lap}{\triangle}

\newcommand{\n}[1]{\label{#1}}

\newcommand{\CAL}{\mathcal}

\newcommand{\ts}[1]{{\boldsymbol{#1}}}

\newcommand{\ST}{ Schwarzschild-Tangherlini \ }

\def\XXint#1#2#3{{\setbox0=\hbox{$#1{#2#3}{\int}$ }
\vcenter{\hbox{$#2#3$ }}\kern-.6\wd0}}

\begin{document}

\title{Quasitopological gravity and double-copy formalism}

\author{Valeri P. Frolov}%
\email[]{vfrolov@ualberta.ca}
\affiliation{Theoretical Physics Institute, Department of Physics,
University of Alberta,\\
Edmonton, Alberta, T6G 2E1, Canada
}


\begin{abstract}
We propose a new approach to the quasitopological theory of gravity based on a modified classical double--copy construction. Focusing on static, spherically symmetric configurations, we show that all vacuum solutions of $D$--dimensional quasitopological gravity can be obtained from an auxiliary non--linear electrodynamics defined in a flat $(D+1)$--dimensional spacetime. The gravitational field equations reduce to an algebraic relation between a primary curvature invariant and the electric field strength, while the remaining dynamics is governed by a Gauss--law constraint for a point--like charge in the auxiliary space. This correspondence provides a transparent interpretation of higher--curvature gravitational interactions in terms of non--linear gauge dynamics and explains the absence of higher--derivative terms in the reduced equations. As illustrative examples, we apply the formalism to Born--Infeld and Hayward--type models, obtaining higher--dimensional regular black--hole solutions with a de~Sitter--like core. Our results extend the scope of the double--copy paradigm beyond Einstein gravity and suggest a unifying framework for a broad class of higher--curvature theories.

\medskip

\hfill {\scriptsize Alberta Thy  4-25}
\end{abstract}

\maketitle

\section{Introduction}

The General Relativity (GR) is the most successful and experimentally verified theory of gravity, having passed numerous tests in both weak- and strong-field regimes. Among its most remarkable predictions are black holes. The direct observation of gravitational waves emitted during the coalescence of two black holes has provided compelling confirmation of their existence in the Universe.

Despite these successes, GR faces a well-known difficulty: vacuum stationary black-hole solutions possess curvature singularities. To cure this ``disease'' of General Relativity, many modified theories of gravity have been proposed. One of the more recent approaches is the so-called quasitopological theory of gravity (QTG). In this model, the Einstein--Hilbert action is supplemented by an infinite set of polynomial scalar curvature invariants. In a $D$-dimensional spacetime the action takes the schematic form
\be
\begin{split}
W_{QTG}&=\dfrac{1}{2\kappa}\int d^D x \sqrt{-g} \ L(g,R)\, ,\\
& L(g,R)=R+\sum_{n=2}^{\infty} \alpha_n \ell^{2(n-1)} \CAL{Z}_n\, .
\end{split}
\ee
Here $\mathcal{Z}_n$ denotes the sum of all scalar invariants of order $n$ constructed from the Riemann curvature tensor, while $\alpha_n$ are dimensionless coupling constants. The parameter $\ell$, which has the dimension of length, plays the role of a fundamental scale: it sets the characteristic curvature scale at which the higher-order terms begin to contribute significantly.

Variation of the action with respect to the metric yields the corresponding Euler--Lagrange equations,
\[
\frac{\delta W_{QTG}}{\delta g^{\mu\nu}} = 0,
\]
which generalize the Einstein equations by including contributions from the quasitopological curvature invariants $\CAL{Z}_n$.
\be\n{EL}
\begin{split}
&P^{\alpha\beta\gamma\delta}  R_{\lambda\beta\gamma\delta}
-\dfrac{1}{2} \delta^{\alpha}_{\lambda}L
-2\nabla^{\beta} \nabla^{\gamma} P^{\alpha}_{\ \ \beta\gamma\lambda}=0\, ,\\
&P^{\alpha\beta\gamma\delta}=\dfrac{\pa L}{\pa R_{\alpha\beta\gamma\delta}}\, .
\end{split}
\ee

In a general case, the last term in this equation contains derivatives of the metric up to fourth order.
It has been shown that, for $D \ge 5$, and for each $n \ge 2$, one can construct a polynomial $\mathcal{Z}_n$ such that the corresponding Euler--Lagrange equations \eqref{EL}, when restricted to a spherically symmetric metric, contain no derivatives higher than second order  \cite{Bueno:2019ltp, Bueno:2022res,Moreno:2023rfl,Bueno:2024dgm}.
Models of this type are referred to as \emph{quasitopological theories of gravity} \cite{Oliva_2010,Myers:2010ru, Bueno:2016cu, Hennigar:2017ego, Bueno:2019ltp, Bueno:2019ycr, Bueno:2022res,Moreno:2023rfl}.
For such theories the spherically reduced action can be written in a closed analytic form.

It has further been demonstrated that, provided the infinite series does not truncate and the coefficients $\alpha_n$ satisfy certain constraints, the solutions of \eqref{EL} describe \emph{regular} (i.e., nonsingular) black holes \cite{Bueno:2024dgm}.
A substantial body of literature has now accumulated on this subject (see, e.g., \cite{Bueno:2024dgm,Frolov:2024hhe,Bueno:2025qjk, Bueno:2025zaj, Ling:2025ncw, Sueto:2025ufn, Bueno:2025gjg, PinedoSoto:2025hel,Hao:2025utc} and references therein).

The goal of this paper is to present a new approach to quasitopological gravity based on the so-called \emph{double-copy formalism}.
This framework relies on the Kerr--Schild representation of the ``curved'' metric $g_{\mu\nu}$,
\be \n{AKS}
g_{\mu\nu}=\eta_{\mu\nu}+2\Phi k_{\mu} k_{\nu}\, .
\ee
Here $\eta_{\mu\nu}=\mbox{diag}(-1,1,\ldots , 1)$ is a flat metric, $k^{\mu}$ is a null vector field and $\Phi$ is a scalar function in the flat spacetime.
 The main idea of this approach is based on the observation that for the metrics which allow the Kerr-Schild representation the non-linear Einstein equations can be reduced to the linear equations for Maxwell and scalar fields. In particular, both Schwarzschild and Kerr metric, can be easily derived in this approach (see e.g. \cite{Visser:2007fj} and references therein). This method can also be applied to wide class of solutions of the Einstein equations \cite{Monteiro:2014cda,Luna:2015paa,White:2017vfl,Bern:2018jmv}.
 This observation can also be used to simplify calculations of gravity scattering amplitudes by reducing this problem to the calculation of the Yang–Mills amplitudes with a subsequent double copy prescription \cite{Bern_2010,Monteiro_2014,Luna_2015,Bah:2019sda}. At the moment there exist dozens of publications on this subject. Related references can be found e.g. in the following review articles \cite{Monteiro:2014cda,Luna:2015paa,White:2017vfl,White_2018,bern2019duality,Bern:2019prr}.

 The main result of the present paper is demonstration that all vacuum spherically symmetric solutions of $D-$dimensional quasitopological theory of gravity can be obtained by solving a specially chosen
non-linear electodynamics equations in a flat $(D+1)-$dimensional spacetime.

The paper is organized as follows. In section~II we remind basic known results of the quasitopological theory of gravity and their spherically symmetric vacuum solutions. We also discuss a double-copy approach to static spherically-symmeric solutions of four and higher dimensional Einstein equations. In section~III we  describe a special modification of the double-copy formalism. This modification is used for study of solutions of the quasitopological theory of gravity. In section~IV we consider special cases of Hayward-type and Born-Infeld models. Section~V contains discussion of the obtained results.

In the paper we use sign conventions adopted in the book \cite{MTW}.

\section{Schwarzschild-Tangherlini metric in the double copy formalism}

\subsection{Schwarzschild-Tangherlini solution}

In this section we discuss Schwarzschild-Tangherlini solution of the higher dimensional Einstein equations and collect  useful relations. This subject is well known and our main goal is to fix notations and normalizations which will be used later in the paper.

Let $\CAL{M}^D$ be $D$-dimensuinal spacetime  with metric $g_{\mu\nu}$
\be
ds^2=g_{\mu\nu} dx^{\mu} dx^{\nu}\hh \mu, \nu=0,1,\ldots D-1\, .
\ee
We write the Einstein-Hilbert action in the form
\be\n{WW}
W_g=\dfrac{1}{2\kappa}\int d^D x \sqrt{-g} R\, ,
\ee
where $R$ is a scalar curvature and
\be
\kappa=8\pi G^{(D)}\, .
\ee
Here $G^{(D)}$ is a $D$-dimensional Newton's coupling constant. The Einstein equations derived from this action are
\be \n{EE}
R_{\mu\nu}-\dfrac{1}{2}R g_{\mu\nu}=\kappa T_{\mu\nu}\, .
\ee

Let us consider a spherically symmetric static metric in $\CAL{M}^D$ of  the form
\be \n{Nf}
ds^2=-N^2 f dt^2+\dfrac{dr^2}{f}+r^2 d\Omega^2_{D-2}\, ,
\ee
where $ d\Omega^2_{D-2}$ is the metric on a unit $(D-2)-$ dimensional unit sphere $S^{D-2}$. The surface area of this sphere is $\Omega_{D-2}$, where
\be
\Omega_{n}=\dfrac{2 \pi^{(n+1)/2}}{\Gamma((n+1)/2}\, .
\ee

Schwarzschild-Tangherlini solution describes the spacetime of a $D-$dimension static spherically symmetric black hole. It
 is a solution of the vacuum Einstein equations which has the form \eqref{Nf} with
\be\n{ET}
N=1\hh f=1-\dfrac{\mu}{r^{D-3}}\, .
\ee
The constant $\mu$ is related with the mass $M$ of the black hole
\be
\mu=\dfrac{2\kappa M}{(D-2)\Omega_{D-2}}\, .
\ee
This metric can be obtained by substituting the ansatz \eqref{Nf} into the equations \eqref{EE} and solving the obtained ordinary differential equations.
Another option is to first derive the reduced action $W_{red}$ by substituting the ansatz \eqref{Nf} into the Einstein-Hilbert action. The variation of the reduced action  $W_{red}$ with respect to its arguments $N$ and $f$ gives a set of reduced equations for this variables. Schwarzschild-Tangherlini metric is a solution of these reduced equations\footnote{
In fact, this result is applicable to Lie symmetry reduction for any group-invariant Lagrangian. The corresponding statement is known as Palais’ Principle of Symmetric Criticality \cite{Palais:1979rca}. Its derivation and applications were discussed in \cite{Fels:2001rv,Anderson:1999cn}; see also \cite{Frausto:2024egp}.}.

Calculating the Ricci scalar for the metric \eqref{Nf} and substituting it into the action \eqref{WW} after integrating by parts and integrating over the angle variables one obtains (for details, see e.g. \cite{Frolov:2024hhe})\footnote{Since the integrand does not depend on time $t$, the integral over $t$ is understood as an interval $\Delta t$ of time between some initial $t_1$ and final $t_2$ moments of time, $\Delta t=t_2-t_1$. This constant factor does not affect the field equations.}
\be \n{WWR}
W_{red}=B\int dt dr N (r^{D-1}p)'\, .
\ee
Here and later a prime denote a derivative over $r$, and
\be
B=\dfrac{(D-2)\Omega_{D-2}}{2\kappa}\, .
\ee
We also denote
\be \n{pp}
p=\dfrac{1-f}{r^2}\, .
\ee
This is one of the four basic curvature invariants of a spherically symmetric static metric (see e.g. \cite{Frolov:2024hhe}). Since it does not contain derivatives of $f$ and play an important role in what follows we call it a primary invariant.
By varying \eqref{WWR} over $f$ and $N$ one gets
\be
N'=0\hh (r^{D-1}p)'=0\, .
\ee
The first equation show that one can put $N=1$, while the second equation implies
\be\n{ppp}
p=\dfrac{\mu}{r^{D-1}}\, .
\ee
Using \eqref{pp} one restore the Schwarzschild-Tangherlini metric \eqref{Nf}--\eqref{ET}.

\subsection{Double-copy form of the Schwarzschild-Tangherlini metric}

In the double-copy approach one uses the Kerr-Schild form of the metric \eqref{AKS}
where $\eta_{\mu\nu}$ is a flat metric in $D-$dimensional Minkowski spacetime $M^D$, $k^{\mu}$ are null  vectors tangent to a shear-free congruence of null geodesics and $\Phi$ is a scalar potential. Let us note that if $\ts{k}$ is null in the flat metric  $\eta_{\mu\nu}$ it is also null in the curved metric $g_{\mu\nu}$.
This double-null property \eqref{AKS} is what gives Kerr–Schild metrics their remarkable integrability properties.

To write the metric \eqref{Nf} with $N=1$ in the Kerr-Schild form one introduces
the advanced time coordinate\footnote{Note that, instead of the advanced time $v$, one may use the  retarded time coordinate $u$, which is defined analogously by replacing the plus sign in \eqref{vvv} with a minus sign.
}
\be\n{vvv}
v=t+\int\dfrac{dr}{f(r)}\, ,
\ee
and define new time coordinate $T=v-r$. Then
the Schwarzschild-Tangherlini metric can be written as follows
\be\n{KS}
ds^2=dS^2+2\Phi dv^2\, .
\ee
Here $dS^2$ is flat metric in Minkowsky spacetime $M^D$. Denote by $X^{\mu} =(T,X_1,\ldots , X_{D-1})$, $\mu=0,1,\ldots,D-1$, Cartesian coordinates in $M^D$ then
\be
\begin{split}
&dS^2=-dT^2+d\vec{X}\cdot d\vec{X}\, ,\\
&d\vec{X}\cdot d\vec{X}
=\sum_{i=1}^{D-1} (dX_i)^2=dr^2+r^2 d\Omega_{D-2}^2
\,  .
\end{split}
\ee
Here and later we use a notation $\vec{X}=(X_1,\ldots, X_{D-1})$ for $(D-1)-$dimensonal vector in a flat $(D-1)-$dimensonal Euclidean space $E^{D-1}$.
Denote by $\ts{k}$ such a vector that
\be \n{kkk}
\begin{split}
k_{\mu}dx^{\mu}&=-dv=-(DT+dr)\, ,\\
 k^{\mu}\pa_{\mu}&=\pa_T-\pa_r\, .
 \end{split}
\ee
This is a  null vector tangent to radial incoming null rays. For the \ST metric one has
\be\n{AAA}
\Phi =\dfrac{\mu}{r^{D-3}}\, .
\ee

Let us remind how solving the linear Maxwell equations in a flat spacetime $M^D$ allows one to "construct" the \ST solution of the Einstein equations.
Consider the $D-$dimensional Maxwell field $F_{\mu\nu}=A_{\nu,\mu}-A_{\nu ,\mu}$
with action
\be
W_A=-\dfrac{1}{4}\int d^D X F_{\mu\nu}F^{\mu\nu}+\int d^D X J^{\mu} A_{\mu}\, .
\ee
We are looking for a static spherically symmetric solution
with vector potential $A_{\mu}$ of the form
\be
A_{\mu}=\Phi(r) k_{\mu}\, .
\ee
Since $\Phi(r)dr$ is a pure gauge, one has
\be
A_{\mu}=-\Phi \delta^T_{\mu}\, .
\ee
The strength field is
\be
F_{\mu\nu}=\Phi_{,\mu} k_{\nu}-\Phi_{,\nu} k_{\mu}\, .
\ee
Hence
\be
F^{\nu}_{\ \mu} k^{\mu}=E k^{\nu}\, ,
\ee
where $E=-\Phi'$ is the  radial electric field.
It is easy to show that the Maxwell equations $F^{\mu\nu}_{\ \ ;\nu}=J^{\mu}$ imply that the potential $\Phi$ satisfies the following $(D-1)-$dimensional Poisson equation
\be
\lap_{D-1}\Phi=J^0\, .
\ee
Let
\be
J^{\mu}=q\delta^{D-1}(\vec{X})\delta^{\mu}_{T}
\ee
be a current of  a point-like charge $q$ located at the origin of the coordinates then one has
\be
\lap_{D-1}\Phi=q\delta^{D-1}(\vec{X})\, .
\ee
A decreasing at the infinity solution of this equation is
\be
\Phi=\dfrac{q}{(D-3)\Omega_{D-2}}\dfrac{1}{r^{D-3}}\, .
\ee
This solution coincides with \eqref{AAA} for
\be
q=(D-3)\Omega_{D-2}\ \mu=\dfrac{2(D-3)}{D-2} \kappa M\, .
\ee

Let us summarize. The Tangherlini--Schwarzschild solution of the nonlinear \(D\)-dimensional Einstein equations can be constructed by solving the following auxiliary problem: find a solution of the linear Maxwell equations in flat $D-$dimensional spacetime $M^{D}$ for a static point-like source. The eigenvectors of the resulting electromagnetic field $F_{\mu\nu}$ form a shear-free congruence of null vectors $k^{\mu}$ tangent to radial null rays. These null vectors, together with the scalar potential $\Phi$ generated by the point-like charge, uniquely determine the required solution \eqref{AKS} of the Einstein equations.

\section{Modified double-copy approach}

\subsection{Modification of the double-copy approach}

In this section we demonstrate how the \ST solution can be obtained by a slightly different method which we call a modified double-copy formalism.  As a starting point we consider again Maxwell theory in a Minkowski spacetime, but now we assume that its dimension differs from $D$ and is equal to $D+1$. To distingush this model from the one considered in the previous section we adopt the following notations. We denote this space by $M^{D+1}$ and the Cartesian coordinates in it by $\CAL{X}^a$. Small Latin indices, such as $a$, $b$, take values $0,1,\ldots, D-1, D$. We also denote \be
T=\CAL{X}_0\hh \vec{\CAL{X}}=(\CAL{X}_1,\ldots ,\CAL{X}_D)\hh
\CAL{R}^2=\vec{\CAL{X}}\cdot \vec{\CAL{X}}\, .
\ee
The line element in $M^{D+1}$ is
\be
d\CAL{S}^2=-dT^2+d \vec{\CAL{X}}\cdot d \vec{\CAL{X}}\, .
\ee

We write the action of the  $(D+1)-$dimensional Maxwell field
\be
 \CAL{F}_{ab}=\CAL{A}_{b ,a}-\CAL{A}_{a  ,b}\, ,
 \ee
as follows
\be\n{WM}
W_\CAL{A}=-\dfrac{1}{4}\int d^{D+1} \CAL{X} \CAL{F}_{ab}\CAL{F}^{ab}+\int d^{D+1} \CAL{X}  \CAL{J}^{a} \CAL{A}_{a}\, .
\ee

We are looking for  a static spherically symmetric solution of
the Maxwell equations
\be \n{MMM}
\CAL{F}^{a b}_{\ \ ;b}=\CAL{J}^{a}\, ,
\ee
for  a point-like charge  $\CAL{Q}$ located at the origin of the coordinates
\be\n{JJJ}
\CAL{J}^{a}=\CAL{Q}\delta^{D}(\vec{\CAL{X}})\delta^{a}_{T}\, .
\ee
Let us write the corresponding vector potential $\CAL{A}_{a}$ in the form\footnote{
Let us note that the term $-\Psi(\CAL{R})d\CAL{R}$ can be "gauged away". After this gauge transformation one gets a standard expression
\be \n{Psi}
\CAL{A}_{a}=-\Psi(\CAL{R}) \delta_a^T\, ,
\ee
for the potential in the Coulomb gauge.
}
\be
\CAL{A}_{a}=-\Psi(\CAL{R}) \CAL{K}_a\hh \CAL{K}_a d\CAL{X}^a=-(dT+d\CAL{R})\, .
\ee
For this potential the field strength is
\be \n{FF}
\CAL{F}_{ab}=\Psi_{,a} \CAL{K}_b-\Psi_{,b} \CAL{K}_a \, .
\ee
Note that a null vector $\CAL{K}^a$ is an eigenvector of $\CAL{F}_{ab}$
\be
\CAL{F}^{a}_{\ b} \CAL{K}^b=\CAL{E}  \CAL{K}^a\hh
\CAL{E}=- \CAL{K}^a\Psi_{,a}=-\dfrac{d\Psi}{d \CAL{R}}\, .
\ee
It is easy to check that
\be
\CAL{E}^2=-\dfrac{1}{2}\CAL{F}_{ab}\CAL{F}^{ab}\, .
\ee

Maxwell equations \eqref{MMM} for the source \eqref{JJJ} can be written in the form
\be
\dfrac{ 1}{\CAL{R}^{D-1}}\dfrac{d }{d \CAL{R}}\Big(
\CAL{R}^{D-1}\CAL{E}
\Big)=\CAL{Q}\delta^{D}(\vec{\CAL{X}})\, .
\ee

Using expression of the $\delta -$function in spherical coordinates
\be
\delta^{D}(\vec{\CAL{X}})=\dfrac{1}{\Omega_{D-1} \CAL{R}^{D-1}} \delta(\CAL{R})\, ,
\ee
one gets
\be \n{EQ}
\dfrac{d }{d \CAL{R}}\Big(
\CAL{R}^{D-1}\CAL{E}
\Big)=\dfrac{\CAL{Q}}{\Omega_{D-1}}  \delta(\CAL{R})\, .
\ee
Hence
\be \n{EEE}
\CAL{E}=\dfrac{\CAL{Q}}{\Omega_{D-1}\CAL{R}^{D-1}}\, .
\ee

In order to obtain the \ST metric by using these results one can proceed as follows.
\begin{itemize}
\item Denote by $M^D$ a $D-$dimensional hyperplane defined by the condition $\CAL{X}^D=0$.
\item Denote by $X_i$ the Cartesian coordinates $\CAL{X}_i$ , $i=1,\ldots ,D-1$ in  $M^D$.
\item For points in  $M^D$ put $r=\CAL{R}$ .
\item Choose $\CAL{Q}=\Omega_{D-1}\mu$\, .
\end{itemize}
Let us denote a restriction $\pi$ of the objects in $M^{D+1}$ on $M^D$ by $(\ldots )|_D$. Then one has
\be \n{ppkk}
\pi: \ \ p=\CAL{E}|_D\hh \ts{\CAL{K}}|_D=\ts{k}\, .
\ee

Let us summarize. The \ST solution of the vacuum $D-$dimensional  Einstein equations can be constructed by using a static spherically symmetric solution of $(D+1)-$dimensional Maxwell equations for a point-like source by applying the above described procedure. The Kerr-Schild form \eqref{KS} of the solution of the equations of the quasitopological theory of gravity is obtained by using the relation
\be \n{rrpp}
2\Phi=r^2 p\, .
\ee
Denote $v=T+r$, then \eqref{KS} takes the form
\be \n{vvpp}
\begin{split}
ds^2&=-f(r) dv^2+2 dv dr+r^2  d\Omega^2_{D-2}\, ,\\
f&=1-r^2 p(r)\, .
\end{split}
\ee
In order to get this solution in $(t,r)-$coordinates one defines the time coordinate $t$ as follows
\be \n{ttTT}
t=T+r-\int\dfrac{dr}{f(r)}\, .
\ee
In these coordinates the solution has the form \eqref{Nf} with $N=1$ and
$f=1-r^2 p(r)$.

\subsection{Modified double-copy approach for quasitopological theory of gravity}

We demonstrate now how this approach can be generalized and applied to finding solutions of the quasitopological  gravity equations. For this purpose, we consider a non-linear electrodynamics in a flat $(D+1)-$dimensional spacetime $M^{D+!}$ with the action of the form
\be\n{NNLM}
\begin{split}
&W_\CAL{A}=\int d^{D+1} \CAL{X} \ L(\CAL{E})+\int d^{D+1} \CAL{X}  \CAL{J}^{a} \CAL{A}_{a}\, ,\\
&\CAL{E}^2=-\dfrac{1}{2} \CAL{F}_{ab}\CAL{F}^{ab}
\, .
\end{split}
\ee
For $L(\CAL{E})=1/2 \CAL{E}^2$ this action coincides with \eqref{WM}.  We assume that expansion of $L(\CAL{E})$ for small $\CAL{E}$ has the form
\be\n{LLL}
L(\CAL{E})\approx\dfrac{1}{2}\CAL{E}^2+\ldots\, ,
\ee
where dots denote higher in $\CAL{E}$ terms.
This condition guarantees that, in the weak-field regime, the theory reduces to the linear Maxwell electrodynamics. Nonlinear electrodynamics is a well-established subject with a vast literature. Comprehensive discussions can be found in the books and review articles \cite{BornInfeld1934,Plebanski1970,
Alam2021,Boillat1970,Dirac1960,Avetisyan2022} , which also contain extensive further references.

For study spherically symmetric solutions of the form \eqref{Psi} it is convenient to use the reduced form of the field action \eqref{NNLM}
\be \n{WRED}
W_{red}[\Psi]=\Omega_{D-1} \int dT   d\CAL{R} \ \CAL{R}^{D-1} h(\CAL{E})\hhh \CAL{E}=-\dfrac{d\Psi}{d\CAL{R}}\, .
\ee
Denote
\be\n{hhEEE}
h(\CAL{E})=\dfrac{dL}{d\CAL{E}}\, .
\ee
Then variation of the action $W_{red}[\Psi]$ with respect to $\Psi$ gives
\be\n{hhRR}
\dfrac{d}{d\CAL{R}}\Big(
\CAL{R}^{D-1} h(\CAL{E})
\Big)=\dfrac{\CAL{Q}}{\Omega_{D-1}} \delta(\CAL{R})\, .
\ee
For $h=\CAL{E}$ this equation correctly reproduces \eqref{EQ}, as it should be.

Let us note, that this result can be obtained from a slightly different form of the reduced action
\eqref{WRED}
\be \n{WN}
\CAL{W}(N,\CAL{E})=-\Omega_{D-1} \int dT d\CAL{R}\ N(\CAL{R})\ \dfrac{d}{d\CAL{R}}\Big( \CAL{R}^{D-1} h(\CAL{E})\Big)\, ,
\ee
where $N(\CAL{R})$ is an arbitrary function playing the role of the Lagrange multiplier. Note, that in the action \eqref{WN} the other argument of this functional is $\CAL{E}$ (not $\Psi$).

Solving \eqref{hhRR} one gets
\be
h=\dfrac{\CAL{Q}}{\Omega_{D-1}\CAL{R}^{D-1}}\, .
\ee
Following the prescription described in the previous section and relation \eqref{ppkk} one finds\footnote{
Let us note that in the original derivation of the reduced action for the quasitopological theory of gravity the condition $D\ge 5$ is usually imposed. In our derivation of this metric we do not impose this condition.
}
\be \n{PROC}
p=\CAL{E}|_D \hhh
h(p)=\dfrac{\mu}{r^{D-1}} \hhh
k_{\mu}dX^{\mu}=-(dT+dr)\, .
\ee

Let us summarize. To obtain a solution of the quasitopological theory of gravity one first solves the equations for a non-linear electroctromagnetic field in $(D+1)-$dimensional flat spacetime. Using \eqref{hhEEE} one finds the function $h(\CAL{E})$ for
the corresponding Lagrangian density. Then, using the projection $\pi$ of $M^{D+1}$ onto $M^D$, and relations \eqref{PROC} one obtains $h=h(p)$ as a function of $r$. Inverting the expression for $h$ one finds $p=p(r)$. After this, following the prescription described at the end of the previous section and using relations \eqref{rrpp}--\eqref{ttTT} one obtains a required solution  the quasitopological theory of gravity in $(t,r)-$coordinates.

\begin{figure}[!htb]%
    \centering
 \includegraphics[width=0.2\textwidth]{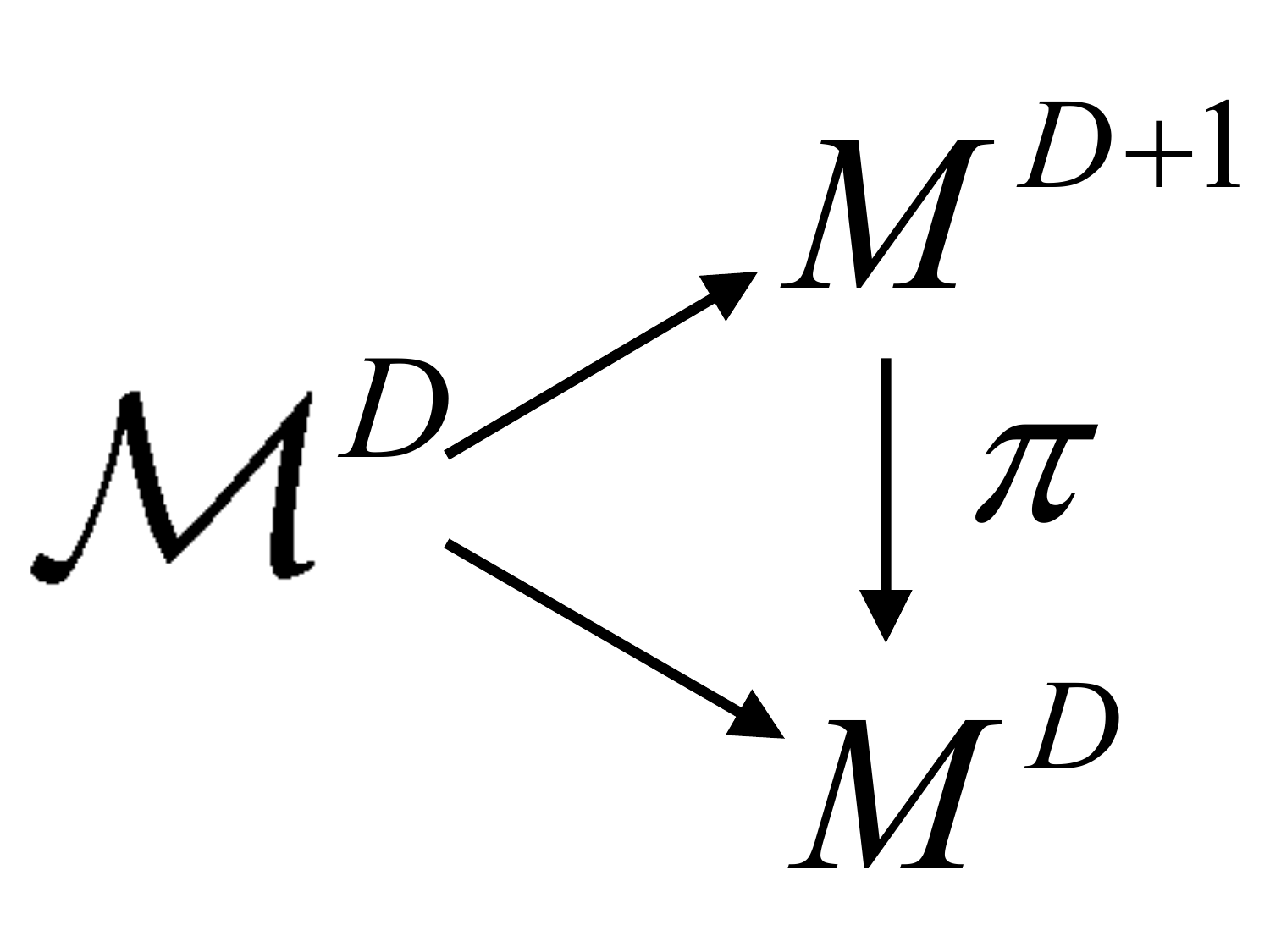}\\[0pt]
\caption{Illustration of the modified double-copy approach.
To obtain a solution of the quasitopological gravity field equations in a curved
\(D\)-dimensional spacetime \(\mathcal{M}^D\), one first solves the equations of a
nonlinear electrodynamics in a \((D+1)\)-dimensional \textit{flat} spacetime
\(M^{D+1}\).
The resulting solution is then restricted to a \(D\)-dimensional subspace
\(M^D\), and this restricted field configuration is used to reconstruct the
curved spacetime metric, written in Kerr--Schild form, via the double-copy
representation.}
    \label{F1}
\end{figure}

\subsection{Scaling properties}

Let us show that solutions obtained in the modified double-copy formalism possess important scaling properties. Denote by $\ell$ the length parameter which enters the non-linear Lagrangian density. Both objects $p$ and $h(p)$ have dimensions of $[length]^{-2}$. We denote dimensionless versions of these objects by
\be
\hat{p}=\ell^2 p\hh \hat{h}=\ell^2 h\, .
\ee
For $N=1$ besides $\hat{p}$ there exist two more dimensionless basic curvature invariants, which we denote by $\hat{q}$ and $\hat{v}$
\be \n{qqvv}
\hat{q}=\hat{p}+\dfrac{1}{2}r\hat{p}'\hh
\hat{v}=\hat{p}+2r\hat{p}'+
\dfrac{1}{2}r^2\hat{p}''\, ,
\ee
(for details, see e.g. \cite{Frolov:2024hhe}).
We also define a dimensionless coordinate $y$
\be
y=\dfrac{r^{D-1}}{\mu \ell^2}\, .
\ee
Then \eqref{qqvv} can be written in the form
\be\n{qqqvvv}
\begin{split}
&\hat{q}=\hat{p}+\dfrac{D-1}{2} \ y\ \pa_y\hat{p}\, ,\\
& \hat{v}=\hat{p}+\dfrac{(D-1)(D+2)}{2}\ y\ \pa_y\hat{p}+
\dfrac{(D-1)^2}{2}\ y^2\ \pa_y^2\hat{p}
\end{split}
\ee
The equation $\hat{h}(\hat{p})=1/y$ defines $\hat{p}$ as a function of $y$, while relations \eqref{qqqvvv} define other dimensionless basic curvature invariants as functions of $y$. Scalar polynomial curvature invariants of the metric \eqref{Nf} with $N=1$ can be written as polynomial functions of $\hat{p}$, $\hat{q}$ and $\hat{v}$.
The condition \eqref{LLL} implies that far from the source
\be
\hat{p}(y)\approx \dfrac{1}{y}\, .
\ee

Let us assume that the Lagrangian density $L(\mathcal{E})$ is chosen such that the corresponding function $\hat{p}(y)$ is finite at $y=0$ and bounded on the semi-axis $y \in [0,\infty)$. Then the relations $\eqref{qqqvvv}$ show that the other basic curvature invariants possess similar properties. As a result, all polynomial curvature invariants are uniformly bounded. This implies that this class of theories satisfies the Markov limiting curvature condition \cite{Markov:1982,Markov:1984ii}.

\section{Special cases}

\subsection{Born-Infeld model}

Born-Infeld model \cite{BornInfeld1934} is one of the best known theory of non-linear electrodynamics\footnote{
For general discussion of the non-linear electrodynamics see \cite{BornInfeld1934,Dirac1960,Plebanski1970,Boillat1970,FairlieGovaerts1992,Alam2021,Avetisyan2022} and references therein.
}. In this section we use this model in the modified double-copy approach and discuss properties of the corresponding black hole solutions in the quasitopological gravity. Our starting point is the following Born-Infeld action in $M^{D+1}$
\be\n{NLM}
\begin{split}
&W_{BI}=-\int d^{D+1} \CAL{X} \ L(\CAL{F}_{ab})\, ,\\
& L(\CAL{F}_{ab})=\dfrac{1}{\ell^4}\Big(
1-\sqrt{1+\dfrac{\ell^4}{2}\CAL{F}_{ab}\CAL{F}^{ab}}
\Big)\, .
\end{split}
\ee
For a weak field one has
\be
L(\CAL{F}_{ab})\approx -\dfrac{1}{4} \CAL{F}_{ab}\CAL{F}^{ab}+\ldots \, .
\ee
Here $\ell$ is a length-scale parameter. A spherically reduced version of the Lagrangian density $L$ is
\be
L(\CAL{E})=\dfrac{1}{\ell^4}\Big( 1-\sqrt{1-\ell^4\CAL{E}^2}
\Big)\, .
\ee
Simple calculation gives
\be \n{hhEE}
h(\CAL{E})\equiv \dfrac{dL}{d\CAL{E}}=\dfrac{\CAL{E}}{\sqrt{1-\ell^4\CAL{E}^2}}\, .
\ee
This relation implies
\be
\CAL{E}=\dfrac{{h}}{\sqrt{1+\ell^4h^2}}\, .
\ee
Using relations \eqref{PROC} one finds
\be
p=\dfrac{\mu}{\sqrt{r^{2(D-1)}+\mu^2 \ell^4}}\, .
\ee
The corresponding metric function in the metric \eqref{Nf} (with $N=1$) is
\be
f=1-\dfrac{\mu r^2}{\sqrt{r^{2(D-1)}+\mu^2 \ell^4}}\, .
\ee
This is a required solution of the spherically reduced equation of the quasitopological gravity theory.

\subsection{Hayward model}

As another interesting example we consider a non-linear electrodynamics in $M^{D+1}$ with the following spherically-reduced Lagrangian density
\be
L=-\dfrac{1}{\ell^4}\Big(
\ell^2 \CAL{E}+\ln (1-\ell^2 \CAL{E})
\Big)\, .
\ee
For small $\CAL{E}$ one has
\be
L=\dfrac{1}{2}\CAL{E}^2+\dfrac{1}{3}\ell^2 \CAL{E}^3\, .
\ee
Simple calculation gives
\be
h=\dfrac{\CAL{E}}{1-\ell^2 \CAL{E}}\, .
\ee
After projection to $M^D$ and using $\CAL{E}|_D=p$ one gets
\be
h(p)=\dfrac{p}{1-\ell^2 p}\, .
\ee
Inversion of this relation gives
\be
p=\dfrac{h}{1+\ell^2 h}\, .
\ee
Using expression $h={\mu}/{r^{D-1}}$,
one obtains the following expression for the metric function
\be
f=1-\dfrac{\mu r^2}{r^{D-1}+\mu \ell^2}\, .
\ee
The corresponding solution of the quasitopological gravity is nothing, but a higher dimensional version of the Hayward metric.

\section{Discussion}

In this work we have demonstrated that the problem of finding static, spherically symmetric vacuum solutions of the $D$--dimensional quasitopological theory of gravity can be reformulated in a remarkably simple and transparent way using a modified double--copy construction. The key observation is that, for metrics admitting a Kerr--Schild representation, the highly nonlinear gravitational field equations reduce to a set of algebraic and ordinary differential relations involving a single ``primary'' curvature invariant $p(r)$. We have shown that these relations can be obtained from an auxiliary non--linear electrodynamics defined in a flat $(D+1)$--dimensional spacetime.

More precisely, the gravitational dynamics encoded in the infinite series of quasitopological curvature invariants is mapped onto the constitutive relation of a non--linear electrodynamics theory, specified by the function $h(\CAL{E})=dL/d\CAL{E}$. The gravitational field equation determining the metric function $f(r)$ is then equivalent to the Gauss law for a point--like electric charge in the auxiliary flat space. This correspondence provides a concrete realization of a classical double--copy--type relation beyond Einstein gravity and highlights the deep structural similarity between quasitopological gravity and non--linear gauge theories.
An important advantage of this approach is its universality. Once the mapping between the gravitational invariant $p(r)$ and the electric field $\CAL{E}(R)$ is established, any choice of non--linear electrodynamics Lagrangian immediately generates a corresponding spherically symmetric solution of the quasitopological gravity equations. In particular, models that are known to regularize the electric field at the origin lead, after the double--copy projection, to black--hole geometries with a de~Sitter--like core and finite curvature invariants. The Born--Infeld and Hayward--type examples considered here illustrate how regular black holes arise naturally in this framework.

From a conceptual point of view, the modified double--copy construction clarifies why quasitopological theories admit regular black--hole solutions without introducing higher--derivative instabilities in the spherically symmetric sector. The absence of higher--order derivatives is reflected, on the gauge--theory side, in the purely algebraic constitutive relation between $\CAL{E}$ and $h(\CAL{E})$. This provides an intuitive explanation of the integrability of the reduced gravitational equations and suggests that other classes of modified gravity theories admitting similar reductions may also possess an underlying gauge--theoretic description.

There are several natural directions for future work. First, it would be interesting to extend the present analysis beyond static and spherically symmetric configurations, for example to rotating solutions or time-dependent backgrounds. Second, the role of matter sources in the gravitational theory—and their interpretation within the auxiliary nonlinear electrodynamics—deserves further investigation. It would also be important to address key issues such as the stability of regular black holes \cite{DeFelice:2025fzv} and the problem of mass inflation near the inner horizon \cite{Carballo-Rubio:2018pdc}. Finally, it would be worthwhile to explore whether the correspondence established here can be embedded into a broader theoretical framework, potentially shedding new light on the inner structure of black holes in quasitopological gravity.
Overall, the results of this paper provide additional evidence that the double--copy paradigm is not restricted to Einstein gravity, but instead represents a more general organizing principle linking gravity and gauge theories, even in the presence of higher--curvature interactions.

\section*{Acknowledgments}
The author thanks the Natural Sciences and Engineering Research Council of Canada and the Killam Trust for their financial support.



%

\end{document}